\documentclass[aip,reprint,showpacs,numerical,twocolumn,preprintnumbers,amsmath,amssymb]{revtex4-1}

\usepackage{graphicx}
\usepackage{dcolumn}
\usepackage{bm}
\usepackage{amssymb}
\usepackage{color}
\usepackage{ulem}

\begin{document}

\title{Lateral vibration effects in atomic-scale friction}

\author{R. Roth}
\affiliation{Climate and Environment Physics, Physics Institute, University of  Bern, Bern, Switzerland}
\affiliation{Oeschger Centre for Climate Change Research, University of Bern, Bern, Switzerland}

\author{O.~Y.~Fajardo}
\affiliation{Departamento de F\'{\i}sica de la Materia Condensada and Instituto  de Ciencia de Materiales de Arag\'{o}n, CSIC-Universidad de  Zaragoza, 50009 Zaragoza, Spain}

\author{J.~J.~Mazo}
\affiliation{Departamento de F\'{\i}sica de la Materia Condensada and Instituto  de Ciencia de Materiales de Arag\'{o}n, CSIC-Universidad de  Zaragoza, 50009 Zaragoza, Spain}

\author{E. Meyer}
\affiliation{Department of Physics, University of Basel, Klingelbergstrasse 82,  4056 Basel, Switzerland}

\author{E. Gnecco}
\affiliation{Instituto Madrile\~{n}o de Estudios Avanzados en Nanociencia,  IMDEA Nanociencia, 28049 Madrid, Spain}

\date{\today}

\begin{abstract}
The influence of lateral vibrations on the stick-slip motion of a
nanotip elastically pulled on a flat crystal surface is studied by
atomic force microscopy (AFM) measurements on a NaCl(001) surface in
ultra-high vacuum. The slippage of the nanotip across the crystal
lattice is anticipated at increasing driving amplitude, similarly to
what is observed in presence of normal vibrations. This lowers the
average friction force, as explained by the Prandtl-Tomlinson model
with lateral vibrations superimposed at finite temperature.
Nevertheless, the peak values of the lateral force, and the total
energy losses, are expected to increase with the excitation amplitude,
which may limit the practical relevance of this effect.
\end{abstract}


\maketitle

Developing strategies for lowering friction is a key issue for proper
functioning of micro- and nano-electromechanical systems (MEMS and
NEMS). In this context, traditional lubricants cannot be used, since
the viscosity of mineral oils dramatically increases when the
lubricant molecules are confined into nanometer-sized
interstices~\cite{tl_hu98}. Different approaches, such as mechanical
excitations, need to be explored.  Ultrasonic vibrations have been
used for years to modify the frictional behavior of materials at a
macroscopic scale~\cite{Akay2002} and their application at the
nanoscale looks quite promising~\cite{apl_dinelli97, jap_behme01,
  SocoliucScience2006, NN_Lantz09, jast_vanspengen10}.  Here, we are
particularly interested in sharp asperities sliding on atomically flat
surfaces.  In this case, exciting mechanical resonances of the
nano-junctions formed while sliding can be a valid method to reduce
friction. This was shown by Socoliuc {\it et
  al.}~\cite{SocoliucScience2006} in atomic-scale AFM experiments on
alkali halide surfaces in ultra-high vacuum (UHV) and by Lantz {\it et
  al.}~\cite{NN_Lantz09}, who succeeded in preventing the abrasive
wear of a silicon tip sliding over several hundred meters following
this strategy.  The state of `dynamic superlubricity' so-achieved was
also exploited to acquire lattice-resolved friction force maps of
crystal surfaces without damaging the samples~\cite{NT_Gnecco09}.
Note that, in the previous experimental works, the actuation was
applied perpendicular to the sliding plane. Friction reduction was
also predicted theoretically when mechanical oscillations are induced
parallel to this plane~\cite{ti_tshiprut07}.  However, although this
effect was observed in a series of macroscopic measurements based on a
tribometer~\cite{Popov10}, the variation of the friction features on
the atomic scale has not been documented so far.

In this work we present AFM measurements in UHV complementing the
results in Ref.~\onlinecite{SocoliucScience2006}, where the effect of
normal excitations was investigated by applying an ac voltage between
the tip and a counter electrode on the backside of an insulating
KBr(001) surface.  Here, lateral vibrations of the probing tip are
induced by shaking a piezo-element attached to the cantilever sensor
of the AFM at a frequency corresponding to the torsional resonance of
the cantilever in contact with the sample surface.  An alkali halide
crystal, i.e. the NaCl(001) surface, was chosen as a model system for
the measurements.  Similarly to the results presented in
Ref.~\onlinecite{SocoliucScience2006}, the average friction on the
probing tip decreases at increasing driving amplitude, which can be
explained with the classical Prandtl-Tomlinson (PT) model for
atomic-scale
friction~\cite{Prandtl28,GneccoPRL2000,SangPRL2001,ReguzzoniPNAS2009,OFajardoPRB2010,prl_jansen10,prb_muser12}.
However, this may not be the case for the peak values of the friction
force, which are numerically expected to increase slightly when
ultralow values of the average friction are approached.

\begin{figure}[bt]
\centering
\includegraphics[scale=1.0]{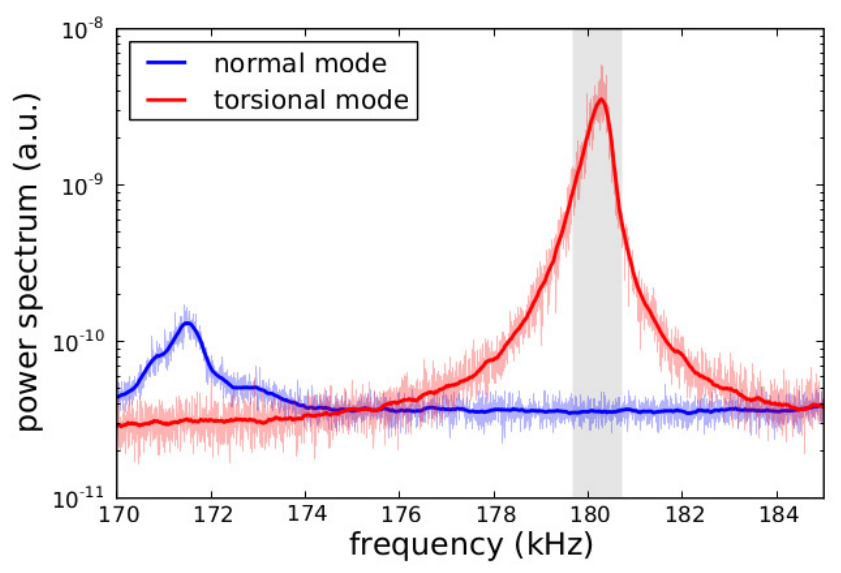}
\caption{Thermal noise power spectrum of the normal
  (blue curve) and torsional vibrations (red curve)  of
    the AFM sensor used in the experiment in contact with a NaCl(001)
    surface in UHV.}
\label{PSD}
\end{figure}

The NaCl(001) sample surface was cleaved in a ultra-high vacuum (UHV)
chamber and analyzed using a home-built AFM~\cite{apl_howald93} and a
silicon tip attached to an elastic cantilever with normal spring
constant $k_N=0.08$~N/m, and torsional spring constant $k_T=53$~N/m
(Nanosensors PPP-CONT).  The cantilever torsion is related to the
(lateral) friction force experienced by the tip by standard formulas
of continuum mechanics~\cite{Bhushan}.  The thermal noise power
spectrum of the torsional vibrations of the cantilever in contact with
the NaCl(001) surface is shown in Fig.~\ref{PSD}.  The resonance
frequency of the system is found at $f_{t0}=180.3$\,kHz.  Note that
another resonance peak $f_{n1}= 171.5$~kHz is observed, corresponding
to the second flexural mode of the cantilever in contact.  The
relative sharpness of the resonance curves (widths at half-maximum of
about 2\,kHz) allows us to exclude any overlap between the two
resonance modes. To get the optimum friction reduction, a phase-locked
loop (PLL) was used to track the torsional resonance peak and excite
the cantilever at the frequency $f_{t0} = 180.3 \pm 0.5$ kHz (gray bar in
Fig.~\ref{PSD}) while scanning the sample surface. This induced
lateral vibrations of the probing tip.

\begin{figure}[tb]
\centering
\includegraphics[scale=1.]{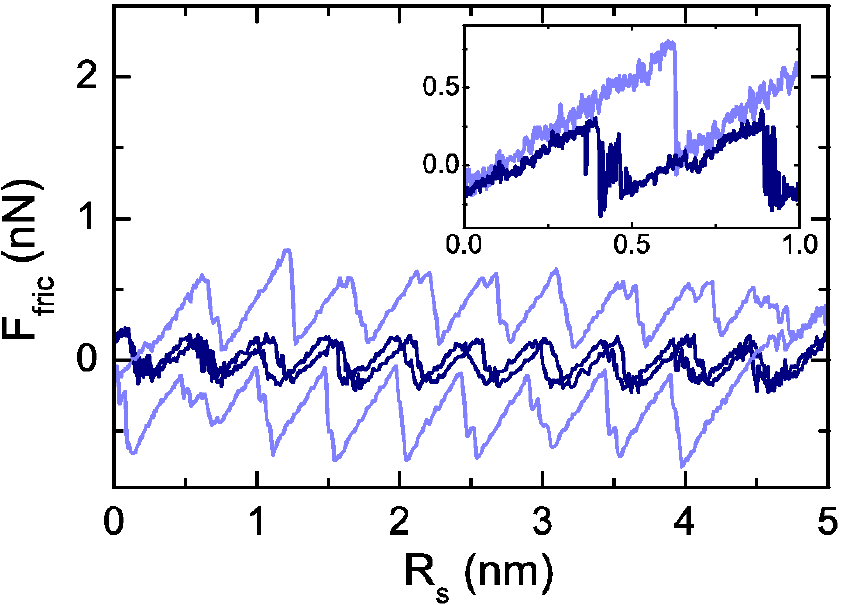}
\caption{Atomic-scale friction force loops measured on NaCl(001) in
  UHV at room temperature with (dark blue) and without (light blue)
  lateral vibrations. $R_s=v_st$ is the cantilever position averaged
  over the vibrations.  The driving amplitude of the $ac$ voltage is
  $U_{exc}=100$\,mV and the frequency $f=180.3$\,kHz corresponds to
  the torsional resonance in Fig. \ref{PSD}.  Each data point is
  averaged over 180 oscillations cycles (over 36 cycles in the inset,
  where only the forward signal is shown).  The normal force and
  velocity values are $F_N=4.9$\,nN and $v_s=10$\,nm/s in both cases.}
\label{fig2}
\end{figure}

\begin{figure}[tb]
\centering
\includegraphics[scale=1.]{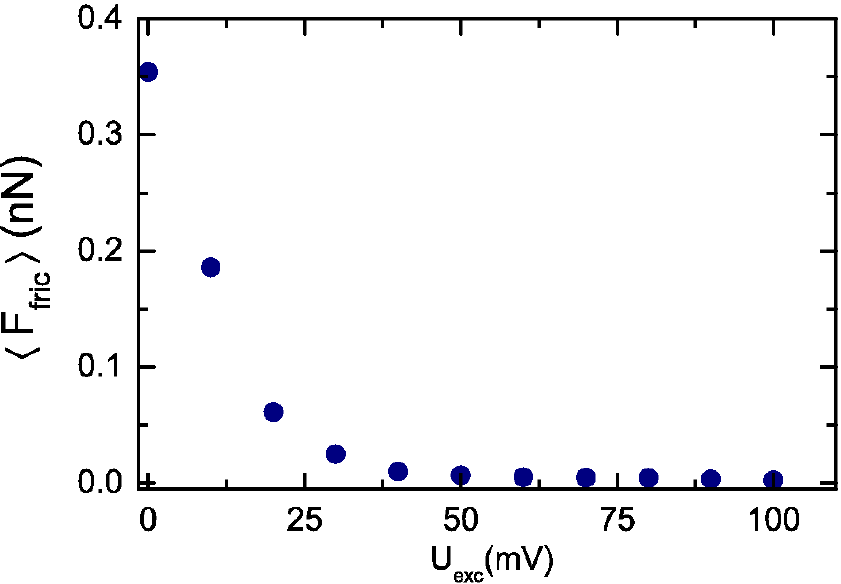}
\caption{Measured average friction force as a function of the amplitude of the
  $ac$ voltage inducing the lateral oscillations.}
\label{fig3}
\end{figure}

Fig. 2 shows two friction loops acquired while shaking the cantilever
at the frequency $f_{t0}$ and without doing that. Note that the
lateral force is significantly smoothed out due to sampling.  In
absence of vibrations, a typical sawtooth pattern is observed. The tip
is pinned to a certain site on the surface lattice till the angle of
torsion of the cantilever reaches a critical value, and the tip
suddenly slips into a nearby location.  When the lateral vibrations
are excited, the sawtooth pattern is shrunk, i.e., the trace and
retrace curves become closer till they completely overlap if the
excitation amplitude is high enough (dark blue curves in
Fig. \ref{fig2}).  In this case the friction force $F_{\rm fric}$; that is, the lateral force averaged over one or multiple oscillation cycles, becomes negligible. Note that the smoothing applied to $F_{\rm fric}$ is much shorter than the washboard frequency. The inset in
Fig.~\ref{fig2} shows also that the complete slip of the tip towards a
new equilibrium position is accompanied by a series of back-and-forth
jumps induced by the thermal vibrations occurring at the finite
temperature ($T=300$ K) of the experiment.  The excitation amplitude
dependence of the friction force, averaged over the stick-slip
movement across the crystal lattice (i.e. of the area enclosed by the
friction loops, divided by the distance travelled by the clamped end
of the cantilever support), is shown in Fig.~\ref{fig3}.  This force
is denoted $\langle F_{\rm fric} \rangle$, to distinguish it from
$F_{\rm fric}$, which is only averaged over a few oscillation cycles. The
force $\left<F_{\rm fric}\right>$ decreases gradually, as observed
when normal oscillations are applied~\cite{SocoliucScience2006}.

\begin{figure}[tb]
\centering
\includegraphics[scale=1.]{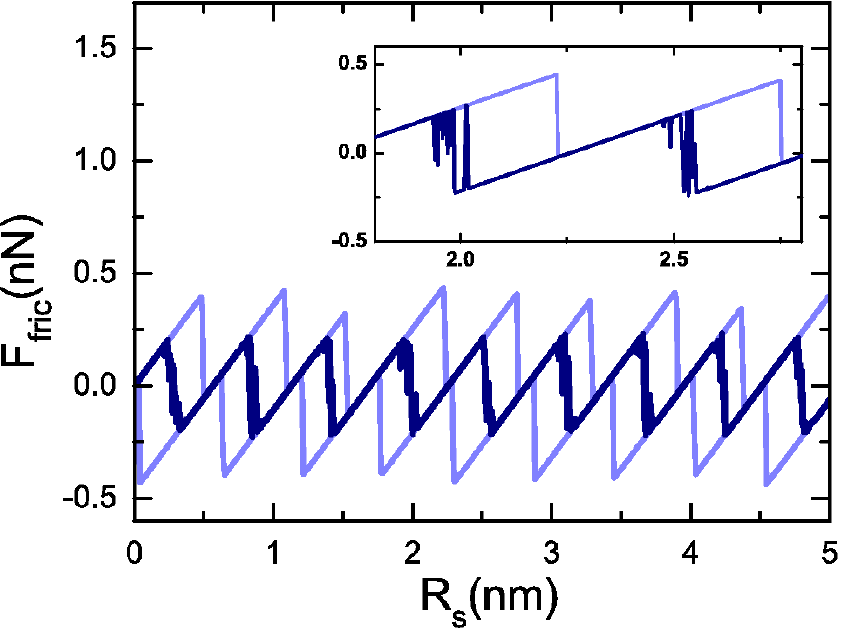}
\caption{Simulated friction force loops with (dark blue) and without
  (light blue) superimposed lateral oscillations.  Parameter values:
  $a=0.564$ nm, $U_0=0.38$ eV, $k=0.95$ N/m, $A=0.6 a$,
  $f=180.3$\,kHz, $v_s=10$\,nm/s, $T=300$\,K. The inset shows the
  back-and-forth jumps accompanying the transition from a lattice site
  to the next one.  As in Fig.~\ref{fig2}, the lateral force is
  averaged over 180 (36 in the inset) oscillation cycles.}
\label{fig4}
\end{figure}

To interpret our results, we have run analytical and numeric
calculations based on the PT model.  Here, a point mass representing
the apex of the probing tip is pulled across a periodic potential of
amplitude $U_0$, simulating the interaction with the surface lattice,
using a spring of stiffness $k$.  The clamped end of the spring moves
with the time $t$ as $R(t)=v_st+A\sin 2\pi ft$, where $A$ and $f$ are
the amplitude and frequency of the lateral vibrations, and $v_s$ is
the scan velocity.  A random noise force $\xi(t)$, which is related to
the temperature $T$ by the fluctuation-dissipation relation, is also
added to the forces acting on the tip, in order to reproduce thermal
effects. A lateral force loop resulting from model values consistent
with the experimental parameters is shown in Fig.~\ref{fig4}. Note
that the value of the effective mass was fixed to $7.7\times 10^{-16}$
kg, corresponding to a resonance frequency of 5.6 MHz and the
simulations were performed at the critical damping.  The average
friction signal is indeed reduced when the lateral vibrations are
included and the trace and retrace signals tend to overlap, as in the
AFM experiments. Back-and-forth jumps are also present, as seen in the
inset of Fig.~\ref{fig4}.

\begin{figure}[tb]
\centering
\includegraphics[scale=1.]{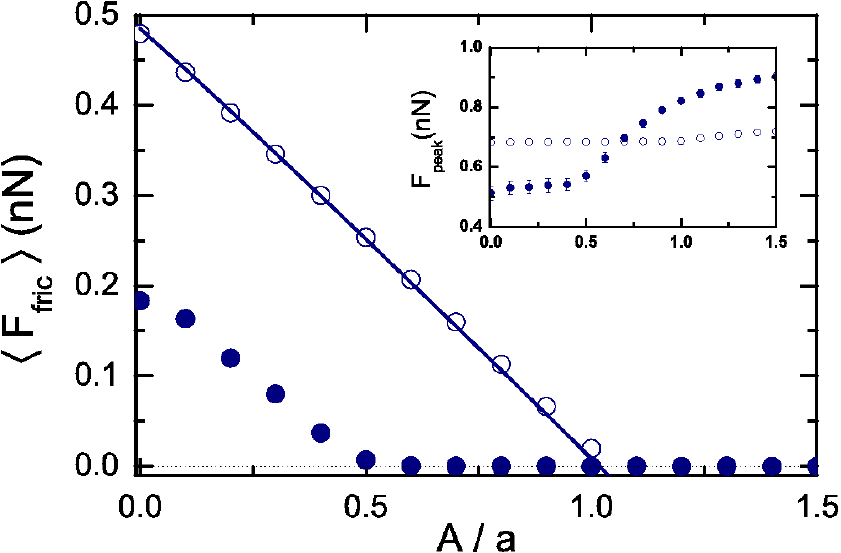}
\caption{Computed lateral force $\left< F_{\rm fric}\right>$, as averaged
  across the crystal lattice, at increasing lateral driving amplitude.
  The results at $T = 0$ K (empty circles) are well-reproduced by the
  continuous line defined by the equation (\ref{equation}).  The
  filled circles are obtained when $T=300$~K.  The corresponding peak
  values of the instantaneous lateral force (not averaged) are shown
  in the inset. We show the mean peak force computed for 100 stick-slip cycles and the obtained error bars.}
\label{fig5}
\end{figure}

\begin{figure}[tb]
\centering
\includegraphics[scale=1.]{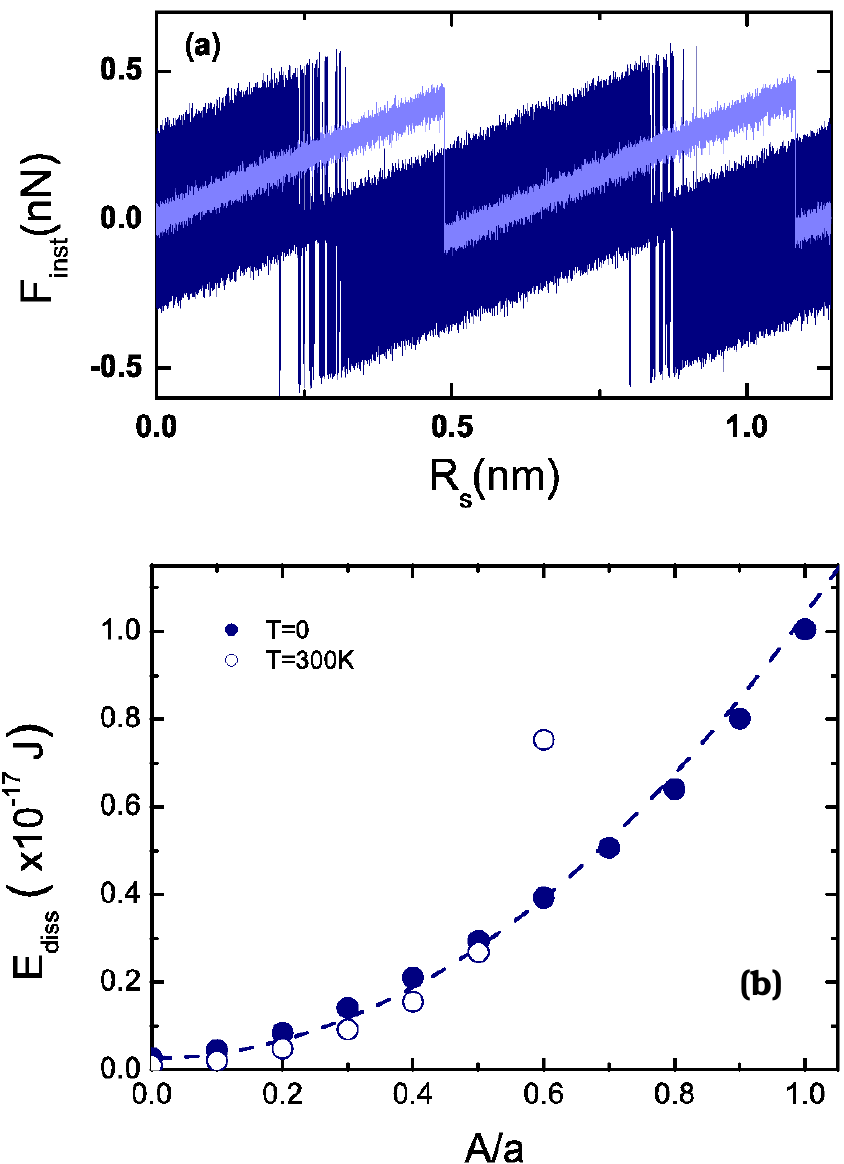}
\caption{(a) Simulated instantaneous variation of the lateral force with (dark
  blue) and without (light blue) superimposed lateral oscillations (the
  parameter values are the same as in Fig.~\ref{fig4}). (b) Energy
  dissipation expected at different values of the excitation amplitude
  $A$ at $T=0$ K (closed circles) and $T=300$ K (open circles). The
  dashed curve shows a quadratic fit expected below the transition to
  the superlubric regime.}
\label{fig6}
\end{figure}

The variation of the average friction force with the
driving amplitude is plotted in Fig. \ref{fig5}, at
both zero and room temperature. We observe that the simulated points
obtained at $T=0$ K are well-fitted by the expression
\begin{eqnarray}
\label{equation}
\left<F_{\rm fric}\right>&(A)& \simeq \frac{ka}{2\pi} \left \{ \eta - \pi
\left(1+\frac{2A}{a}\right) + \right. \nonumber \\ & & \left. +
\frac{4}{3}\sqrt{\frac{\pi }{\eta}} \left[
  \left(1+\frac{A}{a}\right)^{3/2}-\left(\frac{A}{a}\right)^{3/2}
  \right] \right \} ,
\end{eqnarray}
where $\eta = 4\pi^2 U_0/(ka^2)$. Equation~(\ref{equation}) is
obtained by power series expansion, as detailed in
Ref.~\onlinecite{fajardo13}.  As shown in Fig.~\ref{fig5}, it predicts
an almost linear decrease of friction with the oscillation amplitude
in the range of parameters that we considered. This conclusion is
partially modified by the presence of thermal vibrations.  Including
the thermal noise term $\xi(t)$ in the equations of motion, the
$\left<F_{\rm fric}\right>$ vs.~$A$ curve is lowered and slightly
bent, as shown by the filled circles in Fig. \ref{fig5}.  This is not
surprising, since thermal vibrations are also quite effective in
reducing friction on the nanoscale~\cite{SangPRL2001,prl_jansen10}.  A
comparison of Fig.~\ref{fig3} and Fig.~\ref{fig5} shows that an
excitation of 50 mV applied to the piezo actuator leads to lateral
oscillations of approximately half lattice constant of the cantilever
support, suggesting a simple method for a rough calibration of the
driving amplitude.

Our analysis would not be complete without considering the
instantaneous variations of the lateral force $F_\mathrm{inst}(t) =
k[R(t)-x(t)]$, where $R(t)$ and $x(t)$ are the support and tip
position respectively.  As seen in Fig.~\ref{fig6}, this force is
rapidly oscillating and we notice that its peak values do not vary
significantly with the excitation amplitude although the average
lateral force value $\langle F_{\rm fric} \rangle$ decreases.  In the
simulations at room temperature, we even observe an increase of the
peak force when the average force $\langle F_{\rm fric} \rangle$
becomes negligible around $A\simeq 0.5a$ (see inset of
Fig.~\ref{fig5}).  Even more pronounced is the variation of the energy
dissipation $E_{\rm diss}$, which is calculated as the power
$F_\mathrm{inst}\cdot dR/dt$, integrated over a lattice period.  The
energy loss $E_{\rm diss}$ is found to increase with the amplitude
$A$, as shown in Fig.~\ref{fig6}(b).  When the ultralow average
friction regime is entered (when $A\simeq a$ at $T= 0$ or $A\simeq
0.5a$ at $300$ K), the increase of $E_{\rm diss}$ becomes steeper.

In summary, we have performed atomic-scale friction measurements on a
NaCl(001) surface, in which lateral vibrations of the nanoprobe are
excited while sliding on the sample surface. As a result, a continuous
decrease of the average friction force with the driving amplitude is
observed.  The experimental results are supported by numeric
calculations based on the Prandtl-Tomlinson model including thermal
vibrations.  As a drawback, our simulations also predict that,
oppositely to the average friction force, the energy dissipation in
the system may become significant when the probe is vibrated, which
may seriously limit the application of our results to reduce friction
and wear in nanodevices.  A thorough analysis of the lateral force
signal using high bandwidth, as done by Maier et al. in absence of
external excitations~\cite{Maier05}, may help to shed light on this
important issue.  Applications to specific configurations such as
cantilevers oscillating in the pendulum geometry in close proximity to
a solid surface~\cite{Kisiel11, Langer14} are also possible.

\vspace{0.5cm}

We thank Dr. Thilo Glatzel and Dr. Shigeki Kawai for technical
assistance.  O.~Y. F. and J.~J. M. acknowledge financial support from
Spanish MINECO through Project No. FIS2011-25167, cofinanced by FEDER
funds. O.~Y. F. acknowledges financial support from FPU grant by
Ministerio de Ciencia e Innovaci\'on of Spain. E. G. acknowledges
financial support from Spanish MINECO through
Project. No. MAT2012-34487.


\vspace{0.5cm}

\begin{thebibliography}{10}

\bibitem{tl_hu98}
Y. Z. Hu, S. Granick, Trib. Lett. {\bf 5}, 81 (1998).

\bibitem{Akay2002} See, e.g. 
  A. Akay. J. Acoust. Soc. Am. 111, 1525 (2002) and Refs. therein.
  
\bibitem{apl_dinelli97}
F. Dinelli, S. K. Biswas, G. A. D. Briggs, and O. V. Kolosov,
Appl. Phys. Lett. {\bf 71}, 1177 (1997).

\bibitem{jap_behme01}
G. Behme and T. Hesjedal,
J. App. Phys. \textbf{89}, 4850 (2001).

\bibitem{SocoliucScience2006}
A. Socoliuc, E. Gnecco, S. Maier, O. Pfeiffer, A. Baratoff, R. Bennewitz, and E. Meyer, Science {\bf 313}, 207 (2006).

\bibitem{NN_Lantz09} M. A. Lantz, D. Wiesmann, and B. Gotsmann, Nature
  Nanotech. {\bf 4}, 586 (2009).

\bibitem{jast_vanspengen10}
W. M. van Spengen, G. H. C. J. Wijts, V. Turq, and J. W. M. Frenken,
J. Adh. Sci. Tech. \textbf{24}, 2669 (2010).

\bibitem{NT_Gnecco09}
E. Gnecco, A. Socoliuc, S. Maier, J. Gessler, T. Glatzel, A. Baratoff,
and E. Meyer, Nanotechnol. {\bf 20}, 025501 (2009).

\bibitem{ti_tshiprut07}
Z. Tshiprut, A. Filippov, and M.Urbakh, Tribology Int. {\bf 40}, 967 (2007).

\bibitem{Popov10}
V. L. Popov, J. Starcevic, and A. E. Filippov, Tribology Letters {\bf 39}, 25 (2010).

\bibitem{Prandtl28}
L. Prandtl, J. App. Math. Mech. {\bf 8}, 85 (1928). 

\bibitem{GneccoPRL2000}
E. Gnecco, R. Bennewitz, T. Gyalog, C. Loppacher, M. Bammerlin, E. Meyer, and H. J. G{\"u}ntherodt, Phys. Rev. Lett. {\bf 84}, 1172 (2000).

\bibitem{SangPRL2001}
Y. Sang, M. Dube, and M. Grant, Phys. Rev. Lett. {\bf 87}, 174301 (2001).

\bibitem{ReguzzoniPNAS2009}
M. Reguzzoni, M. Ferrario, S. Zapperi, and M. C. Righi,
Proc Natl Acad Sci USA {\bf 107}, 1311 (2009).

\bibitem{OFajardoPRB2010}
O. Y. Fajardo and J. J. Mazo, Phys. Rev. B {\bf 82}, 035435 (2010).

\bibitem{prl_jansen10} L. Jansen, H. H\"{o}lscher, H. Fuchs, and
  A. Schirmeisen, Phys. Rev. Lett. {\bf 104}, 256101 (2010).

\bibitem{prb_muser12}
M. H. M\"{u}ser, Phys. Rev. B {\bf 84}, 125419 (2011).

\bibitem{apl_howald93}
L. Howald, E. Meyer, R. L\"{u}thi, H. Haefke, R. Overney, H. Rudin, and H. J. G\"{u}ntherodt,
Appl. Phys. Lett. \textbf{63}, 117 (1993).

\bibitem{Bhushan} {\it Springer Handbook of Nanotechnology},
  Ed. by B. Bhushan, Springer (2007).

\bibitem{fajardo13}
O. Y. Fajardo, E. Gnecco, and J. J. Mazo, unpublished.

\bibitem{Maier05}
S. Maier, Yi Sang, T. Filleter, M. Grant, R. Bennewitz, E. Gnecco, and
E. Meyer, Phys. Rev. B 72, 245418 (2005).

\bibitem{Kisiel11}
M. Kisiel, E. Gnecco, U. Gysin, L. Marot, S. Rast, and E. Meyer,
Nature Materials {\bf 10}, 119 (2011).
 
 \bibitem{Langer14}
 M. Langer, M. Kisiel, R. Pawlak, F. Pellegrini, G. E. Santoro,
 R. Buzio, A. Gerbi, G. Balakrishnan, A. Baratoff, E. Tosatti, and
 E. Meyer, Nature Materials {\bf 13}, 173 (2014).

\end{thebibliography}
\end{document}